\documentclass[lettersize,journal]{IEEEtran}
\usepackage{amsmath,amsfonts,amssymb}
\usepackage{algorithmic}
\usepackage{algorithm}
\usepackage{array}
\usepackage[caption=false,font=normalsize,labelfont=sf,textfont=sf]{subfig}
\usepackage{textcomp}
\usepackage{stfloats}
\usepackage{relsize}
\usepackage{url}
\usepackage{verbatim}
\usepackage{graphicx}
\usepackage{balance}
\usepackage[colorlinks]{hyperref}
\hypersetup{citecolor=blue}
\usepackage{bm} 

\usepackage{geometry}
\newtheorem{remark}{\bfseries Remark}
\newcommand{\trace}{\mathrm{tr}} 
\newcommand{\Tr}{\trace}
\DeclareMathOperator{\ReOp}{Re}
\newcommand{\E}{\mathbb{E}}

\providecommand{\diag}{diag} 
\renewcommand{\diag}{\mathrm{diag}}

\newcommand{\snr}{\gamma}

\usepackage[font=small,skip=2pt]{caption}

\usepackage{enumitem}
\setlist{nosep}
\begin{document}

\title{\Large {\color{blue}Cram\'{e}r-Rao Bounds for Activity Detection in Conventional and Fluid Antenna Systems}}

\author{Zhentian Zhang,~\IEEEmembership{Graduate Student Member,~IEEE}, 
            Kai-Kit Wong,~\IEEEmembership{Fellow,~IEEE},\\ 
            Hao Jiang,~\IEEEmembership{Senior Member,~IEEE}, 
            Christos Masouros,~\IEEEmembership{Fellow,~IEEE}, and 
            Chan-Byoung Chae,~\IEEEmembership{Fellow,~IEEE}
\vspace{-8mm}

\thanks{Z. Zhang and H. Jiang are with the National Mobile Communications Research Laboratory, Southeast University, Nanjing, 210096, China and H.~Jiang is also with the School of Artificial Intelligence, Nanjing University of Information Science and Technology, Nanjing 210044, China. (e-mail: zhentianzhangzzt@gmail.com, jianghao@nuist.edu.cn).}
\thanks{K. K. Wong and C. Masouros are with the Department of Electronic and Electrical Engineering, University College London, Torrington Place, U.K. (e-mails: \{kai-kit.wong, c.masouros\}@ucl.ac.uk). K. K. Wong is also affiliated with the Yonsei Frontier Lab., Yonsei University, Seoul, 03722 South Korea.}
\thanks{C.-B. Chae is with the School of Integrated Technology, Yonsei University, Seoul, 03722 South Korea (e-mail: cbchae@yonsei.ac.kr).}

\thanks{Corresponding authors: H. Jiang (jianghao@nuist.edu.cn)}}

\maketitle

\begin{abstract}
In this letter, we develop a unified Cram\'{e}r-Rao bound (CRB) framework to characterize the fundamental performance limits of transmission activity detection in fluid antenna systems (FASs) and conventional multiple fixed-position antenna (FPA) systems. To facilitate CRB analysis applicable to activity indicators, we relax the binary activity states to continuous parameters, thereby aligning the bound-based evaluation with practical threshold-based detection decisions. Closed-form CRB expressions are derived for two representative detection formulations, namely covariance-oriented and coherent models. Moreover, for single-antenna FASs, we obtain a closed-form coherent CRB by leveraging random matrix theory. The results demonstrate that CRB-based analysis provides a tractable and informative benchmark for evaluating activity detection across architectures and detection schemes, and further reveal that FASs can deliver strong spatial-diversity gains with significantly reduced complexity.
\end{abstract}


\vspace{-2mm}
\section{Introduction}
\IEEEPARstart{A}{ctivity} detection is one of the essential tasks in systems with sporadic user transmissions \cite{Liu2018,Cov2021}. Two prevailing models are usually considered, namely, \emph{coherent detection} \cite{Liu2018,AMP1,AMP2} and \emph{covariance-based detection} \cite{Cov2021,Cov1}. The former is typically formulated as a Gaussian linear regression problem under white noise, exploiting instantaneous phase and amplitude, whereas the latter relies on second-order statistics and enables activity detection without explicit channel knowledge.

The fundamental limits of both models were characterized in \cite{Cov2021}. Coherent detection requires the blocklength to scale linearly with the number of active users to ensure reliable recovery, while covariance-based detection reduces this requirement to logarithmic order by leveraging large antenna arrays. Despite this, both detectors infer user activity through \emph{parameter relaxation}, i.e., the binary activity indicator is treated as a continuous variable and thresholded. 
{\em This relaxation motivates a unified performance analysis via estimation-theoretic tools.}

Usually, a single-antenna receiver suffers from limited detection performance due to the lack of spatial diversity. Fortunately, fluid antenna systems (FASs) \cite{FAS1} challenge this view by enabling spatial diversity through antenna and electromagnetic reconfiguration with minimal hardware complexity \cite{FAS_tutorial,hong2025contemporary,new2025flar,wu2024flu,Lu-2025}. Hardware implementation techniques for FAS were discussed in \cite{tong2025designs}. Though there were studies in FAS focusing on unsourced massive access \cite{FAS2,FAS3} and finite-blocklength transmission \cite{block_model}, {\em activity detection with FAS is unknown.}

In this letter, we unify the activity detection performance of different systems within the CRB framework for evaluation of performance limits under {\color{blue}quasi-static fading} for FAS and fixed-position antenna (FPA) systems. Our main contributions are as follows:
\begin{itemize}
\item We derive the CRB for covariance-based activity detection in conventional multi-FPA systems, covering both orthogonal (closed-form) and non-orthogonal cases.
\item We derive the coherent CRB for single-antenna FAS and conventional multi-FPA systems, and obtain closed-form approximations via random matrix theory.
\end{itemize}

\textit{Notations}---Scalars, vectors, and matrices are denoted by $x$, $\mathbf{x}$, and $\mathbf{X}$. $(\cdot)^T$, $(\cdot)^H$, and $(\cdot)^*$ denote transpose, Hermitian, and conjugate. $\mathbb{E}[\cdot]$ and $\trace(\cdot)$ denote expectation and trace. $\mathbf{I}_N$ is the identity matrix. $\mathcal{CN}(\mu,\sigma^2)$ denotes a circularly symmetric complex Gaussian distribution, and $\ReOp(\cdot)$ gets the real part.

\vspace{-2mm}
\section{Covariance-Based CRB}\label{sec.2}
In this section, the CRB of the covariance-based activity detector in conventional multi-FPA system is derived.

\subsubsection{\textbf{Signal Model for $M$-FPA Systems}}
Consider an uplink system with $M$ receive FPAs and $K$ potential single-FPA users. Each user is assigned a pilot sequence $\mathbf{s}_k \in \mathbb{C}^{L \times 1}$, and the received signal $\mathbf{Y} \in \mathbb{C}^{M \times L}$ is given by
\begin{equation}\label{eq.1}
\mathbf{Y} = \mathbf{H}\mathbf{B}\mathbf{S} + \mathbf{Z},
\end{equation}
where $\mathbf{H} \in \mathbb{C}^{M \times K}$ is the channel matrix with independent and identically distributed (i.i.d.) entries $h_{m,k} \sim \mathcal{CN}(0, \sigma_h^2)$, $\mathbf{S} = [\mathbf{s}_1, \dots, \mathbf{s}_K]^T\in\mathbb{C}^{K\times L}$ is the pilot matrix with elements drawn from $\mathcal{CN}(0, \bar{p})$, and $\mathbf{Z}$ is the additive white Gaussian noise matrix with zero mean and variance $\sigma_z^2$. The transmission signal-to-noise ratio (SNR) is defined as $\snr = \frac{\bar{p}}{\sigma_z^2}$. User activity is modeled by the diagonal matrix $\mathbf{B} = \diag(b_1, \dots, b_K)$, where in practice $b_k \in \{0,1\}$ indicates whether user $k$ is active. Since the CRB is defined only for continuous, differentiable parameter spaces, we relax the model by treating $b_k \in \mathbb{R}$ as a continuous and deterministic unknown amplitude coefficient.

{\color{blue}The parameter relaxation enables the calculation of the Fisher Information Matrix (FIM) as a function of the {\em signal strength} which is a lower bound of $|b_k|$ which is then converted into the CRB of $b_k$, serving as a benchmark for estimating the continuous activity-strength parameter, particularly at the point $b_k=1$.} Assuming large array size, the empirical covariance matrix $\mathbf{R}$ of the received signal at each antenna is \cite{Cov2021}
\begin{align}
\mathbf{R}(\boldsymbol{\theta}) &= \E[\mathbf{y}_m \mathbf{y}_m^H] \approx \sum_{k=1}^K \theta_k \mathbf{s}_k \mathbf{s}_k^H + \sigma_z^2 \mathbf{I}_L,
\end{align}
where the off-diagonal elements in $\mathbf{R}(\boldsymbol{\theta})$ approach zeros and $\theta_k \triangleq \sigma_h^2 b_k^2$ denotes the effective power strength of the $k$-th user.

\subsubsection{\textbf{Benchmark 1---Covariance-Based CRB}}
To understand how the $M$ FPAs contribute to the total information, we derive the FIM from the joint likelihood. Let $\mathbf{y}_m\in \mathbb{C}^L \times 1{}$ denote the signal received at the $m$-th antenna. Since the channels and noise are independent across antennas, the observations ${\mathbf{y}_m}$, $m\in[1:M]$ are also independent complex Gaussian vectors, $\mathbf{y}_m \sim \mathcal{CN}(\mathbf{0}, \mathbf{R})$. Thus, the joint PDF is written as
\begin{equation}
p(\mathbf{Y}; \mathbf{\theta}) = \prod_{m=1}^M \frac{1}{\pi^L \det(\mathbf{R})} \exp\left( -\mathbf{y}_m^H \mathbf{R}^{-1} \mathbf{y}_m \right).
\end{equation}
The log-likelihood function $\mathcal{L}(\mathbf{\theta})$ can then be obtained as
\begin{equation}
\mathcal{L}(\mathbf{\theta}) = \sum_{m=1}^M \left[ -L \ln \pi - \ln \det(\mathbf{R}) - \mathbf{y}_m^H \mathbf{R}^{-1} \mathbf{y}_m \right].
\end{equation}
The FIM, $\mathbf{J}_{\text{total}}$, is defined as the negative expectation of the Hessian matrix, which is given by
\begin{equation}
\mathbf{J}_{\text{total}} = -\E \left[ \frac{\partial^2 \mathcal{L}(\mathbf{\theta})}{\partial \mathbf{\theta} \partial \mathbf{\theta}^T} \right] = \sum_{m=1}^M \left( -\E \left[ \frac{\partial^2 \ln p(\mathbf{y}_m; \mathbf{\theta})}{\partial \mathbf{\theta} \partial \mathbf{\theta}^T} \right] \right).
\end{equation}
Considering the i.i.d.~assumption among $M$ receiving FPAs, the expectation is identical for each antenna. Thus, the total Fisher information can be feasibly scaled from the Fisher information at single antenna, i.e., $\mathbf{J}_{\text{total}} \leq M \times \mathbf{J}_{\text{single}}$.

Using the Slepian-Bangs formula \cite{Kay1993}, the $(i,j)$-th element of $\mathbf{J}_{\text{single}}$ is $\Tr(\mathbf{R}^{-1} \frac{\partial \mathbf{R}}{\partial \theta_i} \mathbf{R}^{-1} \frac{\partial \mathbf{R}}{\partial \theta_j})$. Since $\frac{\partial \mathbf{R}}{\partial \theta_i} = \mathbf{s}_i \mathbf{s}_i^H$, we have
\begin{align}
[\mathbf{J}_{\mathbf{\theta}}]_{ij} &= M \Tr\left( \mathbf{R}^{-1} \mathbf{s}_i \mathbf{s}_i^H \mathbf{R}^{-1} \mathbf{s}_j \mathbf{s}_j^H \right)
= M \left| \mathbf{s}_i^H \mathbf{R}^{-1} \mathbf{s}_j \right|^2.
\end{align}

In the general scenario where pilots are {\em non-orthogonal}, the FIM is a dense matrix. To further compute the CRB for user $k$, one should first construct the $K \times K$ matrix $\mathbf{J}_{\mathbf{\theta}}$ by computing the quadratic forms $\mathbf{s}_i^H \mathbf{R}^{-1} \mathbf{s}_j$ for all user pairs, then invert the matrix to find $[\mathbf{J}_{\mathbf{\theta}}^{-1}]_{kk}$. Finally, we transform the bound from the power domain $\theta_k$ to the amplitude domain $b_k$ using the Jacobian $\frac{\partial \theta_k}{\partial b_k} = 2 \sigma_h^2 b_k$, yielding
\begin{equation}
\text{CRB}(b_k) = \frac{1}{4 \sigma_h^4 b_k^2} [\mathbf{J}_{\mathbf{\theta}}^{-1}]_{kk},
\end{equation}
which denotes the {\em lower-bound}\footnote{This lower-bound depicts the activity parameters $b_k$ regardless of the users that are not active, which is the best estimation performance for activity.} for the detection on the active users in a conventional multi-FPA system.

In the special case of {\em orthogonal} pilots, i.e., $\mathbf{S}^H \mathbf{S} = L \bar{p} \mathbf{I}$, $\mathbf{R}$ becomes a diagonalized matrix. Using the Woodbury identity \cite{Stoica2005}, the bound simplifies to
\begin{equation}
\text{CRB}(b_k) \approx \frac{(L \bar{p} \sigma_h^2 b_k^2 + \sigma_z^2)^2}{4 M L^2 \bar{p}^2 \sigma_h^4 b_k^2},
\end{equation}
which is proved in Appendix~\ref{appedx.1}.

Moreover, when the system is overloaded, e.g., $K > L$ or $L$ is not sufficiently large, the pilots are inevitably non-orthogonal and consequently, the off-diagonal elements of $\mathbf{J}_{\mathbf{\theta}}$ represent the information loss due to multiuser interference.

\vspace{-2mm}
\section{Coherent CRBs}\label{sec.3}	
In this part, the coherent CRBs for single-antenna FAS and conventional multi-FPA systems are analyzed to demonstrate that a {\em single-antenna FAS} achieves performance comparable to a conventional multi-antenna system with dozens of independent FPAs, implying great simplicity for network design.
\subsubsection{\textbf{Coherent Detection Model for Single-Antenna FAS}}
Conditioned on a specific channel realization $\mathbf{g} \in \mathbb{C}^{1 \times K}$, the single-snapshot observation $\mathbf{y} \in \mathbb{C}^{L \times 1}$ is
\begin{equation}\label{eq.linear_model}
\mathbf{y} = \sum_{k=1}^K b_k (g_k \mathbf{s}_k) + \mathbf{z} = \mathbf{\Phi} \mathbf{b} + \mathbf{z},
\end{equation}
where $\mathbf{s}_k$ denotes the $k$-th column of the pilot matrix $\mathbf{S}^T$, and $\mathbf{\phi}_k = g_k \mathbf{a}_k$ is the effective spatial-temporal signature. The parameter vector of interest is $\mathbf{b} = [b_1, \dots, b_K]^T \in \mathbb{R}^K$. As before, we assume the active set is known and $b_k$ is relaxed. 

\begin{figure*}[t!]
\normalsize
\begin{equation}\label{eq:pdf_channel_response}
\footnotesize
\begin{aligned}
f_{|g_k|^2}(t)
=\sum_{b=1}^B\left\{\int_0^\infty\frac{e^{-\frac{r_b}{2}}L_b}{2}\left[F_{\chi_2^{\prime2}}(\frac{t}{1-\mu^2};\frac{\mu^2r_b}{1-\mu^2})\right]^{L_b-1}\frac{f_{\chi_2^{\prime2}}(\frac{t}{1-\mu^2};\frac{\mu^2r_b}{1-\mu^2})}{1-\mu^2}\mathrm{d}r_b\right\}
\times\prod_{j\neq b}^B\left\{\int_0^\infty\frac{e^{-\frac{r_j}{2}}}{2}\left[F_{\chi_2^{\prime2}}\left(\frac{t}{1-\mu^2};\frac{\mu^2r_j}{1-\mu^2}\right)\right]^{L_j}\mathrm{d}r_j\right\},
\end{aligned}
\end{equation}
\hrulefill
\end{figure*}

\begin{remark}
{\em Unlike Section \ref{sec.2}, which relies on statistical convergence across $M$ antennas, this section considers coherent detection. This model is necessary when $M$ is small or activity must be detected from a single snapshot. In this regime, the sample covariance matrix $\frac{1}{M}\mathbf{Y}\mathbf{Y}^H$ is rank-deficient with rank at most $M$, and when $M \ll K$, it lacks sufficient degrees of freedom to resolve the $K$ user powers.}
\end{remark}

As a consequence, the covariance-based method in Section \ref{sec.2} fails to support the single antenna FAS model since coherent detection requires the phase and amplitude information of the instantaneous received signal. Therefore, coherent-based model is more suitable for single-antenna FASs.

\subsubsection{\textbf{FAS Channel Response Model}}
We employ the spatial block-correlation channel model in \cite{block_model,Block-Correlation}, which offers a favorable tradeoff between modeling accuracy and analytical tractability. Let $\mathbf{g}_k \in \mathbb{C}^{N}$ denote the channel vector of user $k$, with spatial correlation matrix $\mathbf{\Sigma} \in \mathbb{C}^{N \times N}$. Under Clarke's model with uniformly spaced ports, $\mathbf{\Sigma}$ is Toeplitz, i.e.,
\begin{equation}\label{eq.Clarke}
\mathbf{\Sigma}=
\begin{pmatrix}
a(0)&a(1)&\cdots&a(N-1)\\
a(-1)&a(0)&\cdots&a(N-2)\\
\vdots&\vdots&\ddots\\
a(-N+1)&a(-N+2)&\cdots&a(0)
\end{pmatrix},
\end{equation}
where $a(n)=\operatorname{sinc}\!\left(\frac{2\pi nW}{N-1}\right)$ and $W$ is the normalized array length. Spatially correlated channels are generated using the eigenvalue-based construction \cite[(4)]{block_model}
\begin{equation}\label{eq:channel_eigenvalue}
\mathbf{g}_k=\mathbf{Q}\mathbf{\Lambda}^{1/2}\mathbf{g}_{0},
\end{equation}
with $\mathbf{\Sigma}=\mathbf{Q}\mathbf{\Lambda}\mathbf{Q}^{\mathrm{H}}$ and $\mathbf{g}_{0}\sim\mathcal{CN}(\mathbf{0},\sigma_h^2\mathbf{I})$. Although suitable for simulation, direct analytical characterization from the full Toeplitz matrix is intractable.

As observed in \cite{FAS_tutorial}, the spatial correlation is dominated by a few eigenmodes. Accordingly, $\mathbf{\Sigma}$ is approximated by a block-diagonal matrix
\begin{equation}\label{eq:block_correlation_matrix}
\widehat{\mathbf{\Sigma}}=
\begin{pmatrix}
\mathbf{A}_1 & \mathbf{0} & \cdots & \mathbf{0} \\
\mathbf{0} & \mathbf{A}_2 & \cdots & \mathbf{0} \\
\vdots & & \ddots & \vdots \\
\mathbf{0} & \mathbf{0} & \mathbf{0} & \mathbf{A}_B
\end{pmatrix},
\end{equation}
where $\mathbf{A}_b\in\mathbb{R}^{L_b\times L_b}$, $\sum_{b=1}^{B}L_b=N$, and each block follows the constant-correlation model \cite[(20)]{Block-Correlation}
\begin{equation}\label{eq.single_block}
\mathbf{A}_b=\begin{pmatrix}
1 & \mu_b^2 & \cdots & \mu_b^2 \\
\mu_b^2 & 1 & \cdots & \mu_b^2 \\
\vdots & & \ddots & \vdots\\
\mu_b^2 & \cdots & \mu_b^2 & 1
\end{pmatrix},
\end{equation}
with the intra-block correlation coefficient $\mu_b^2\in(0.95,0.99)$. This approximation preserves the essential spatial behavior of \eqref{eq.Clarke} through $\{L_b,\mu_b^2\}$ while enabling tractable statistical analysis. In particular, the PDF of the channel response $|g_k|^2=\max\{|g_{k,1}|^2,\ldots,|g_{k,N}|^2\}$ derived in \cite{block_model} is recalled in \eqref{eq:pdf_channel_response}, where $f_{\chi_2^{\prime 2}}(\cdot;\lambda)$ and $F_{\chi_2^{\prime 2}}(\cdot;\lambda)$ denote the PDF and CDF of a non-central chi-square distribution with two degrees of freedom and non-centrality parameter $\lambda$.

\subsubsection{\textbf{Universal Coherent CRB for FAS}}
The FIM for the linear Gaussian model in \eqref{eq.linear_model}, i.e., $\mathbf{y} \sim \mathcal{CN}(\mathbf{\Phi}\mathbf{b}, \sigma_z^2 \mathbf{I})$ with real parameters $\mathbf{b}$ is given by \cite{Kay1993}
\begin{equation}
[\mathbf{J}_{\mathbf{b}}]_{ij} = \frac{2}{\sigma_z^2} \ReOp\left( \mathbf{\phi}_i^H \mathbf{\phi}_j \right) = \frac{2}{\sigma_z^2} \ReOp\left( g_i^* g_j \mathbf{a}_i^H \mathbf{a}_j \right).
\end{equation}
To derive the CRB for the $k$-th user, we partition the FIM as
\begin{equation}
\mathbf{J}_{\mathbf{b}} = 
\begin{bmatrix} J_{kk} & \mathbf{v}_k^T \\ \mathbf{v}_k & \mathbf{J}_{-k} 
\end{bmatrix},
\end{equation}
where $J_{kk} = \frac{2}{\sigma_z^2} |g_k|^2 \|\mathbf{a}_k\|^2$ is the information from user $k$ alone, and let $\mathbf{v}_k$ represent the cross-information with other users. Using the block matrix inversion lemma, the $k$-th diagonal element of the inverse is given by
\begin{equation}
[\mathbf{J}_{\mathbf{b}}^{-1}]_{kk} = \frac{1}{J_{kk} - \mathbf{v}_k^T \mathbf{J}_{-k}^{-1} \mathbf{v}_k} = \frac{1}{J_{kk}} \cdot \frac{1}{1 - \rho_k^2},
\end{equation}
where $\rho_k^2 = \mathbf{v}_k^T \mathbf{J}_{-k}^{-1} \mathbf{v}_k / J_{kk}$ is the interference factor. Since $\|\mathbf{a}_k\|^2 \approx L \bar{p}$ for Gaussian pilots, we obtain the universal activity detection CRB for FAS as
\begin{equation}\label{eq.universal_CRB_fas}
\text{CRB}(b_k) = \frac{\sigma_z^2}{2 L \bar{p} |g_k|^2} \cdot \frac{1}{1 - \rho_k^2},
\end{equation}
where the factor $\rho_k^2$ depends on the realization of the pilots. 
 
To quantify the impact of multiuser interference, we derive a closed-form approximation of the interference factor $\rho_k^2$ using random matrix theory arguments \cite{Tulino2004} under large blocklength $L$. The quantity $\rho_k^2=\mathbf{v}_k^T\mathbf{J}_{-k}^{-1}\mathbf{v}_k/J_{kk}$ equals the squared cosine of the principal angle between the $k$-th effective signature $\boldsymbol{\phi}_k$ and the interference subspace $\mathcal{S}_{-k}=\mathrm{span}(\{\boldsymbol{\phi}_j\}_{j\neq k})$, i.e.,
\begin{equation}
\rho_k^2=\frac{\left\|\mathbf{P}_{\mathcal{S}_{-k}}\boldsymbol{\phi}_k\right\|^2}{\left\|\boldsymbol{\phi}_k\right\|^2},
\end{equation}
where $\mathbf{P}_{\mathcal{S}_{-k}}$ denotes the orthogonal projector onto the $(K\!-\!1)$-dimensional subspace.

With $\mathbf{S}\sim\mathcal{CN}(0,\bar{p}\mathbf{I})$, the vectors $\{\boldsymbol{\phi}_k\}$ are isotropic in $\mathbb{C}^L$. Hence, the normalized projection energy onto a $(K-1)$-dimensional subspace follows a Beta distribution
\begin{equation}
\rho_k^2\sim\mathrm{Beta}(\alpha,\beta),\quad \alpha=K-1,\ \beta=L-K+1.
\end{equation}
The CRB in \eqref{eq.universal_CRB_fas} contains the penalty $\xi=(1-\rho_k^2)^{-1}$. Let $X=\rho_k^2$ and $Y=1-X$, then $Y\sim\mathrm{Beta}(\beta,\alpha)$. Using $\E[Y^{-1}]=\frac{a+b-1}{a-1}$ for $Y\sim\mathrm{Beta}(a,b)$ with $a>1$, we obtain
\begin{align}\label{eq.expect}
\E\!\left[\frac{1}{1-\rho_k^2}\right]
&=\frac{(L-K+1)+(K-1)-1}{(L-K+1)-1}
=\frac{L-1}{L-K}.
\end{align}
This matches the inverse spectral-efficiency loss in \cite{Tse1999}. Substituting \eqref{eq.expect} into \eqref{eq.universal_CRB_fas} yields the CRB averaged over $\rho_k^2$
\begin{equation}
\E_{\rho_k^2}[\mathrm{CRB}(b_k)] \approx \frac{\sigma_z^2}{2L\bar{p}|g_k|^2}\left(\frac{L-1}{L-K}\right).
\end{equation}
The expression reveals the phase transition in \cite{Chen2019}: as $K\to L$, the factor $(L-K)^{-1}$ diverges, and thus the CRB blows up.

\begin{remark}
 {\em This singularity marks the information-theoretic limit, i.e., no unbiased estimator can reliably recover user activity when the number of active users exceeds the dimension of the pilot space in a single-snapshot coherent system.}
\end{remark}

In conclusion, the CRB averaged over channel fading can be calculated by
\begin{equation}
\E_{|g_k|^2}[\text{CRB}(b_k)] \approx \int_{t=0}^{\infty}\frac{\sigma_z^2}{2 L \bar{p} t} \left( \frac{L - 1}{L - K} \right)f_{|g_k|^2}(t)\mathrm{d}t,
\end{equation}
where $t$ is the random variable denoting channel response  $|g_k|^2$.

\subsubsection{\textbf{Benchmark 2---Coherent CRB for Conventional System}}
To offer fair comparison, we derive the coherent CRB for a conventional multi-FPA system. The received signal $\mathbf{Y} \in \mathbb{C}^{M \times L}$ is modeled as
\begin{equation}
\mathbf{Y} = b_k \mathbf{h}_k \mathbf{s}_k^T + \sum_{j \neq k} b_j \mathbf{h}_j \mathbf{s}_j^T + \mathbf{Z}.
\end{equation}
To eliminate the interference from other $K-1$ users, we construct a projection matrix based on the interference subspace. Let $\mathbf{S}_{\sim k} = [\mathbf{s}_1, \dots, \mathbf{s}_{k-1}, \mathbf{s}_{k+1}, \dots, \mathbf{s}_K] \in \mathbb{C}^{L \times (K-1)}$ be the matrix containing the pilot sequences of the interfering users. The orthogonal projection matrix $\mathbf{P}_{\perp}$ is calculated as
\begin{equation}
\mathbf{P}_{\perp} = \mathbf{I}_L - \mathbf{S}_{\sim k} \left( \mathbf{S}_{\sim k}^H \mathbf{S}_{\sim k} \right)^{-1} \mathbf{S}_{\sim k}^H.
\end{equation}
Multiplying the received signal by $\mathbf{P}_{\perp}$ yields $\mathbf{Y} \mathbf{P}_{\perp}$. Since $\mathbf{s}_j^T \mathbf{P}_{\perp} = \mathbf{0}$ for all $j \neq k$, this yields the effective signal model
\begin{equation}\label{eq.linear_nulling}
\mathbf{Y}_{\text{eff}} = b_k \mathbf{h}_k (\mathbf{s}_k^T \mathbf{P}_{\perp}) + \mathbf{Z} \mathbf{P}_{\perp}.
\end{equation}
For a linear Gaussian model with parameter $\theta$, effective signal vector $\mathbf{u}$, and noise variance $\sigma_z^2$, the Fisher information is defined as $J = \frac{2}{\sigma_z^2} \|\mathbf{u}\|^2$ \cite{Stoica1989}. Here, the parameter is $b_k$ and the effective signal vector is $\mathbf{u} = \mathbf{h}_k (\mathbf{s}_k^T \mathbf{P}_{\perp})$. The conditional FIM can be derived as
\begin{equation}\label{eq.con_FIM}
J_{k|\mathbf{h}_k} = \frac{2}{\sigma_z^2} \|\mathbf{h}_k (\mathbf{s}_k^T \mathbf{P}_{\perp})\|^2 = \frac{2}{\sigma_z^2} \|\mathbf{h}_k\|^2 \|\mathbf{s}_k^T \mathbf{P}_{\perp}\|^2.
\end{equation}

The term $\|\mathbf{s}_k^T \mathbf{P}_{\perp}\|^2$ represents the energy of the target pilot projected onto the subspace orthogonal to the interference. The dimension of the full pilot space is $L$, and the interference subspace has dimension of $K-1$. Thus, the projection preserves $L-(K-1)$ degrees of freedom. The expected effective energy can be easily found as
\begin{equation}\label{eq.exp_energy}
\mathbb{E} \left[ \|\mathbf{s}_k^T \mathbf{P}_{\perp}\|^2 \right] = \bar{p} L \cdot \frac{L-K+1 - 1}{L-1} \approx \bar{p} L \frac{L-K}{L-1}.
\end{equation}
Plugging \eqref{eq.exp_energy} into \eqref{eq.con_FIM}, we obtain
\begin{equation}\label{eq.JJ}
J_{k|\mathbf{h}_k} = \frac{2 \bar{p} L}{\sigma_z^2} \left( \frac{L-K}{L-1} \right) \|\mathbf{h}_k\|^2.
\end{equation}
Also, the unconditional CRB is $\overline{\text{CRB}} = \mathbb{E}_{\mathbf{h}_k}[J_{k|\mathbf{h}_k}^{-1}]$. This requires calculating $\mathbb{E}[\|\mathbf{h}_k\|^{-2}]$. Let $X = \|\mathbf{h}_k\|^2$. Under Rayleigh fading, $X$ follows a Gamma distribution $X \sim \Gamma(M, \sigma_h^2)$ with PDF $f(x) = \frac{1}{\Gamma(M)(\sigma_h^2)^M} x^{M-1} e^{-\frac{x}{\sigma_h^2}}$. The expectation of the inverse variable $1/X$ is derived by integration as
\begin{align}
\mathbb{E}\left[\frac{1}{X}\right] &= \int_{0}^{\infty} \frac{1}{x} \cdot \frac{x^{M-1} e^{-\frac{x}{\sigma_h^2}}}{\Gamma(M)(\sigma_h^2)^M} dx \nonumber \\
&= \frac{1}{(M-1)\Gamma(M-1)(\sigma_h^2)^M} \int_{0}^{\infty} x^{(M-1)-1} e^{-\frac{x}{\sigma_h^2}} dx.
\end{align}
The integral term corresponds to $\Gamma(M-1)(\sigma_h^2)^{M-1}$. Thus,
\begin{equation}\label{eq.exp_Gamma}
\mathbb{E}\left[\frac{1}{X}\right] = \frac{\Gamma(M-1)(\sigma_h^2)^{M-1}}{(M-1)\Gamma(M-1)(\sigma_h^2)^M} = \frac{1}{(M-1)\sigma_h^2}.
\end{equation}
Plugging \eqref{eq.exp_Gamma} into \eqref{eq.JJ}, we obtain the closed-form CRB
\begin{equation}\label{eq:mimo_coh_crb}
\overline{\text{CRB}} = \underbrace{\frac{\sigma_z^2}{2 \bar{p} L}}_{\text{SNR}} \cdot \underbrace{\left(\frac{L-1}{L-K}\right)}_{\text{Interference}} \cdot \underbrace{\frac{1}{(M-1)\sigma_h^2}}_{\text{Array Gain}}.
\end{equation}

\begin{figure}[]
\centering
\includegraphics[width=\columnwidth]{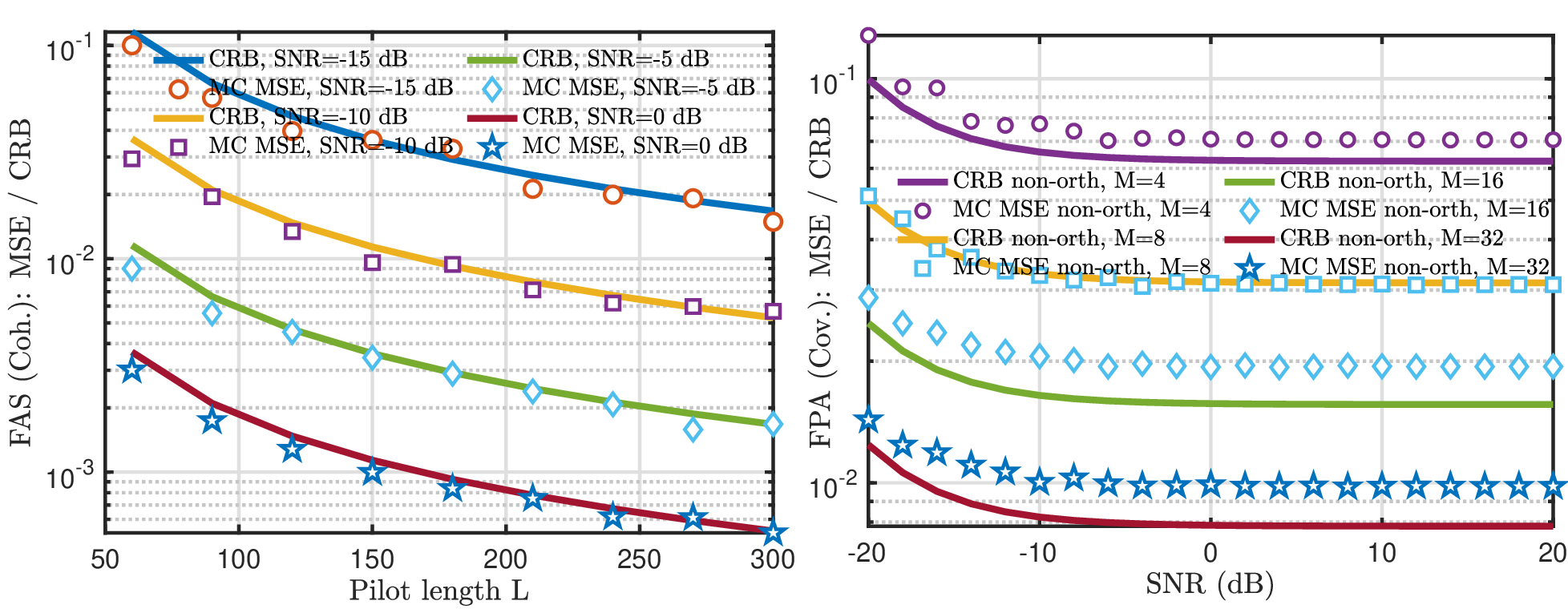}
\caption{Empirical (MSE) vs CRB: $K=20$. For FAS (left), $N=200$ and $W=2$. For covariance-based $M$-FPA (right), $L=400$.}\label{Theory_MC}
\end{figure}

\begin{figure}[]
\centering
\includegraphics[width=.8\columnwidth]{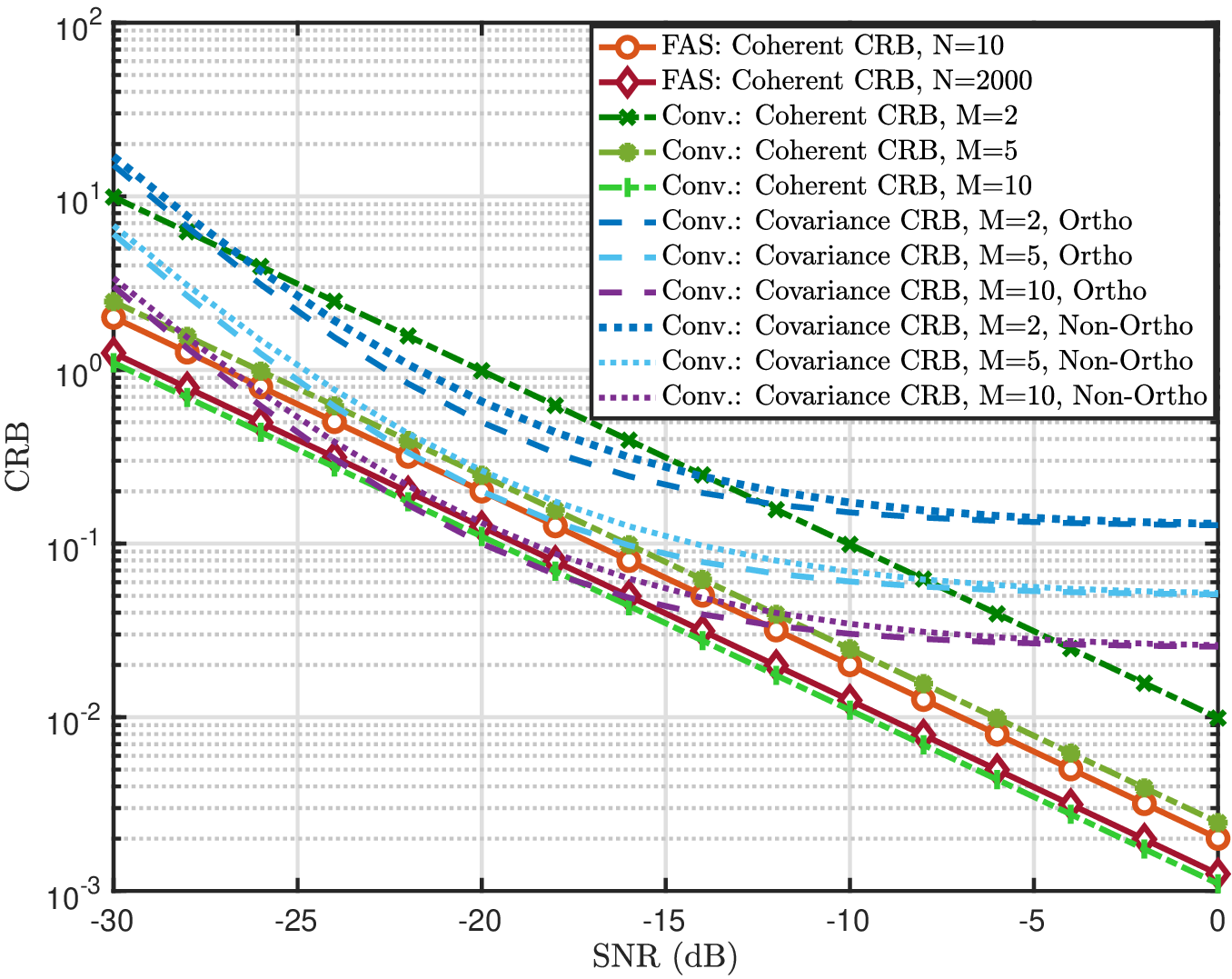}
\caption{CRB versus SNR with $W=5$, $N\in\{10,2000\}$, and $M\in\{2,5,10\}$.}\label{CRB_SNR}
\end{figure}

\vspace{-2mm}
\section{Numerical Results}
Here, we compare the CRBs of coherent and covariance-based schemes and highlight the advantage of FAS over $M$-FPA systems. The block-correlation channel follows \cite{block_model,Block-Correlation} with $\mu=0.97$ and eigenvalue threshold $0.001$. Unless otherwise stated, $L=100$, $K=50$, $\sigma_h^2=1$ and $b_k=1$. 

\subsubsection{Empirical vs. Analysis}
Fig.~\ref{Theory_MC} validates the CRBs by Monte Carlo (MC) mean squared errors (MSEs) of the corresponding practical unbiased detectors. For {\em coherent FAS}, conditioned on the known matrix $\mathbf\Phi$, we estimate $\mathbf b\in\mathbb R^K$ using the unbiased least square estimator $\hat{\mathbf b}=\left(\Re\left(\mathbf\Phi^H\mathbf\Phi\right)\right)^{-1}\Re\left(\mathbf\Phi^H\mathbf y\right)$ and compute $\mathrm{MSE}=\mathbb E[(\hat b_{k_0}-b)^2]$ over independent channel/pilot/noise trials. For short pilots a mild mismatch appears due to strong Gram-matrix fluctuations. For {\em covariance-based} non-orthogonal detection, we exploit second-order moment matching. From $\mathbb E[\mathbf Y\mathbf Y^H]=\mathbf S\mathrm{diag}(\boldsymbol\theta)\mathbf S^H+\sigma_z^2\mathbf I$, taking $\mathbf t=\mathrm{diag}(\mathbf S^H(\hat{\mathbf R}-\sigma_z^2\mathbf I)\mathbf S)$ yields $\mathbb E[\mathbf t]=\mathbf B\boldsymbol\theta$, where $\mathbf B=|\mathbf S^H\mathbf S|^{\circ2}$ denotes the elementwise squared magnitude of the pilot correlation matrix. Thus, the moment-based estimator $\hat{\boldsymbol\theta}=\mathbf B^{-1}\mathbf t$ is {\em unbiased}, and $\hat b=\sqrt{\hat\theta/\sigma_h^2}$. Increasing $M$ reduces MSE, while the gap to the CRB grows for large $M$ as this moment estimator is not efficient.

\subsubsection{CRB vs. SNR}
Fig.~\ref{CRB_SNR} plots the CRBs versus SNR. FAS with $W=5$ and $N\in\{10,2000\}$ consistently outperforms both coherent and covariance-based detectors using $M\in\{2,5,10\}$ antennas across all SNRs. In particular, FAS with $N=2000$ approaches the CRB of a $10$-antenna conventional system and achieves nearly a $10$~dB gain over the $M=2$ case.


\subsubsection{CRB vs. Number of Available Ports $N$}
Fig.~\ref{fig:CRB_N} shows the CRB of FAS versus $N$ under different $W$ at $\mathrm{SNR}=-15$ dB, compared with conventional systems using $M\in\{6,8,12\}$ FPAs. The results indicate that FAS's spatial diversity depends jointly on $W$ and $N$ and is fully exploited only when $N$ is large enough. With $W=5,~10$ and $N=800$, FAS outperforms the covariance-based detector of a $12$-FPA system and approaches its coherent CRB, despite requiring a smaller effective aperture than the half-wavelength-spaced counterpart ($12$ independent antennas require at least $W=5.5$).


\subsubsection{CRB vs. Number of Active Users $K$}
\begin{figure}[]
\centering
\subfloat[]{%
\includegraphics[width=0.8\columnwidth]{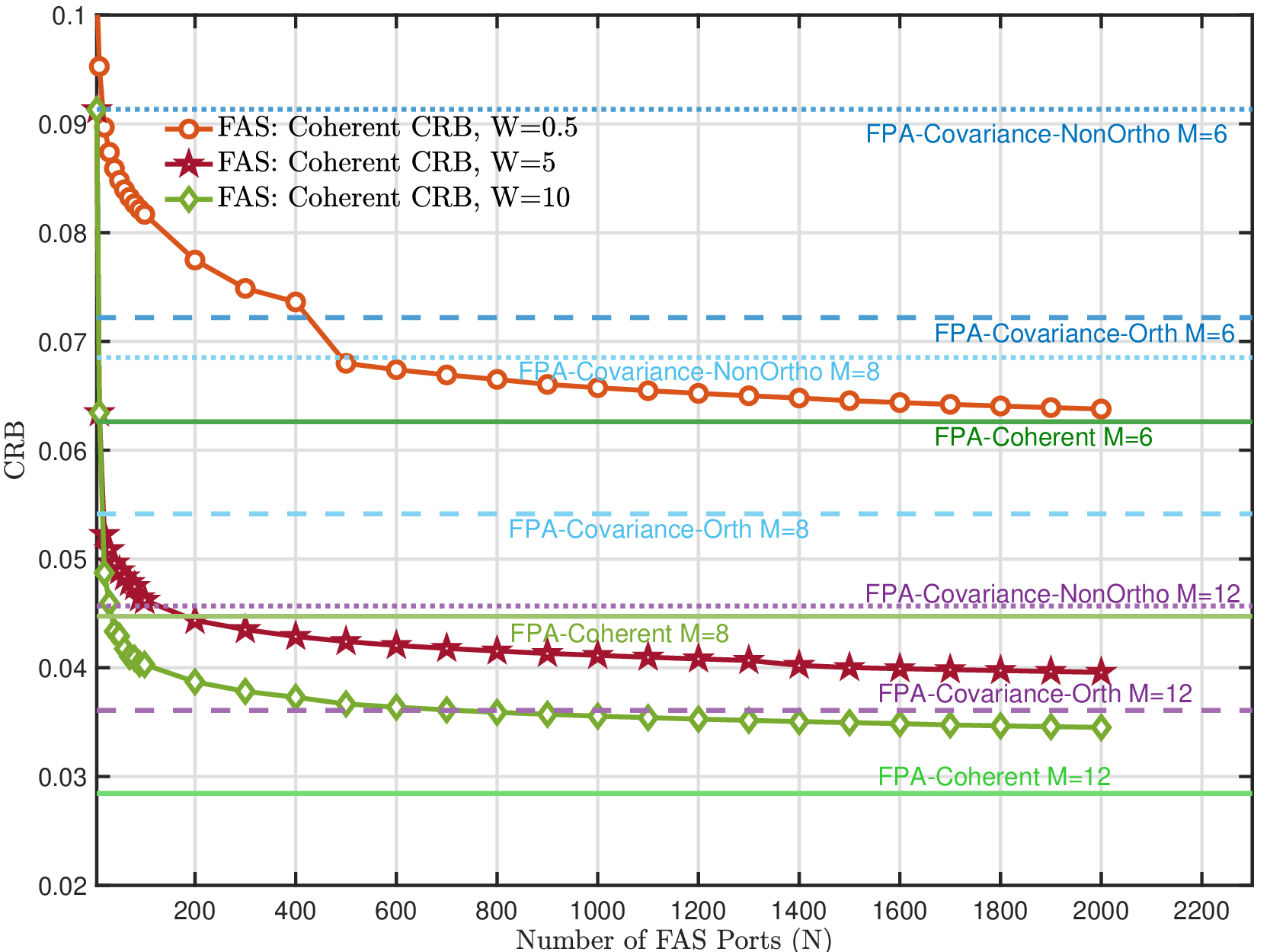}\label{fig:CRB_N}
}\\\vspace{-2mm}
\subfloat[]{%
\includegraphics[width=0.8\columnwidth]{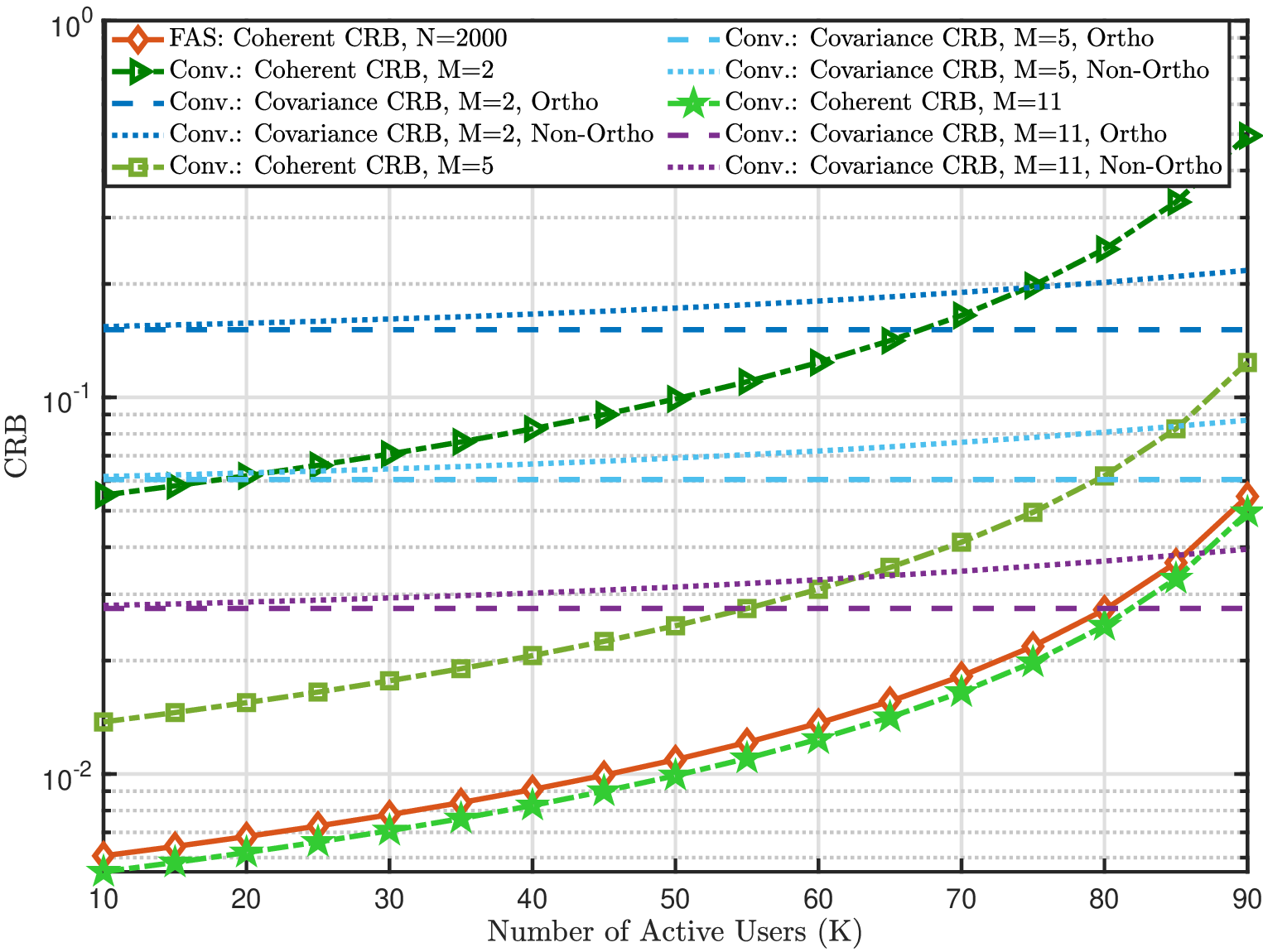}\label{fig:CRB_K}
}
\caption{CRBs of FAS and multi-FPA under different models {\em (a)} CRB versus number of available ports $N$ with $W\in\{0.5,5,10\}$, $M\in\{6,8,12\}$, and $\mathrm{SNR}=-15$ dB; {\em (b)} CRB versus number of active users $K$ with $W=10$, $N=2000$, $M\in\{2,5,11\}$, and $\mathrm{SNR}=-10$ dB.}\label{fig:MSE_Upper}
\end{figure}

Fig.~\ref{fig:CRB_K} depicts the CRB versus $K$ at $\mathrm{SNR}=-10$ dB. Covariance-based detectors exhibit the strongest robustness to access density: the CRB is constant in the orthogonal case since interference is ideally omitted and grows most slowly in the non-orthogonal case, consistent with the logarithmic blocklength scaling in $K$ \cite{Cov2021}. Overall, FAS achieves the lowest CRB and approaches the coherent CRB of a conventional system with $11$ FPAs.

\vspace{-2mm}
\section{Conclusion}\label{sec.5}
{\color{blue}This paper established unified CRB frameworks for AD, with closed-form bounds derived for covariance-based and coherent conventional receivers and for single-antenna FAS via random matrix theory, characterizing fundamental estimation limits and the superiority via spatial diversities from FAS.}


\vspace{-2mm}
\begin{appendices}
\section{Proof on Orthogonal Covariance Detector}\label{appedx.1}
In the special case of orthogonal pilots, we assume $\mathbf{s}_i^H \mathbf{s}_j = 0$ for $i \neq j$ and $\mathbf{s}_i^H \mathbf{s}_i = L \bar{p}$. Under orthogonality, the pilots $\mathbf{s}_k$ are the eigenvectors of $\mathbf{R}$. For any user signatures $\mathbf{s}_i$, we have
\begin{align}
\mathbf{R} \mathbf{s}_i 
&= \theta_i \mathbf{s}_i (\mathbf{s}_i^H \mathbf{s}_i) + \sigma_z^2 \mathbf{s}_i = (\theta_i L \bar{p} + \sigma_z^2) \mathbf{s}_i.
\end{align}
From the eigenvalue property above, it follows that $	\mathbf{R}^{-1} \mathbf{s}_i = \frac{1}{\theta_i L \bar{p} + \sigma_z^2} \mathbf{s}_i$. Since $\mathbf{s}_i^H \mathbf{R}^{-1} \mathbf{s}_j = 0$ for $i \neq j$, the FIM, $\mathbf{J}_{\mathbf{\theta}}$, is diagonal, and the $i$-th diagonal element is given by
\begin{align}
[\mathbf{J}_{\mathbf{\theta}}]_{ii} &= M | \mathbf{s}_i^H \mathbf{R}^{-1} \mathbf{s}_i |^2 = M \left| \frac{\mathbf{s}_i^H \mathbf{s}_i}{\theta_i L \bar{p} + \sigma_z^2} \right|^2 = \frac{M L^2 \bar{p}^2}{(\theta_i L \bar{p} + \sigma_z^2)^2}.
\end{align}
The CRB for power is $\text{CRB}(\theta_i) = [\mathbf{J}_{\mathbf{\theta}}]_{ii}^{-1}$. To find $\text{CRB}(b_i)$, we use $\theta_i = \sigma_h^2 b_i^2$ and get
\begin{align}
\text{CRB}(b_i) &= \left( \frac{\partial \theta_i}{\partial b_i} \right)^{-2} \text{CRB}(\theta_i) 
= \frac{(\sigma_h^2 b_i^2 L \bar{p} + \sigma_z^2)^2}{4 M L^2 \bar{p}^2 \sigma_h^4 b_i^2},
\end{align}
which completes the proof.
\end{appendices}

\vspace{-2mm}
\bibliographystyle{IEEEtran}
\balance

\end{document}